\documentclass{nature}
\usepackage{latexsym}
\usepackage{amsthm}
\usepackage{amssymb}
\usepackage{amsmath}
\usepackage[all]{xy}
\usepackage{graphicx}
\usepackage{subfigure}
\usepackage{dcolumn}
\usepackage{bm}
\usepackage{bbm}
\usepackage{color}
\usepackage{amsfonts}
\usepackage{cases}
\usepackage{float}
\usepackage[a4paper,pagebackref=true,colorlinks=true,
linkcolor=blue,citecolor=blue,
pdfauthor={ },
pdftitle={ },
pdfsubject={ },
pdfkeywords={ }]{hyperref}

\title{Experimental test of non-classicality of quantum mechanics using  an individual atomic solid-state quantum system}

\author{Xi Kong$^{1*}$, Mingjun Shi$^{1*}$, Mengqi Wang$^{1}$\thanks{These authors contributed equally to this work.}, Fazhan Shi$^{1,2}$, Pengfei Wang$^{1}$, Fei Kong$^{1}$, Pu Huang$^{1,2}$, Qi Zhang$^{1}$, Wenchao Ma$^{1}$, Hongwei Chen$^{1}$, Chenyong Ju$^{1,2}$, Mingliang Tian$^{1}$, Changkui Duan$^{1}$, Sixia Yu$^{1}$, \& Jiangfeng Du$^{1,2}$\thanks{djf@ustc.edu.cn}}

\begin{document}
\global\long\def\s{\sigma}
 \global\long\def\k{\kappa}
 \global\long\def\bra#1{\left\langle #1\right|}
 \global\long\def\ket#1{\left|#1\right\rangle }

\maketitle

\begin{affiliations}
 \item National Laboratory for Physical Sciences at the Microscale and Department of Modern Physics, University of Science and Technology of China, Hefei, 230026, China
 \item Synergetic Innovation Center of Quantum Information and Quantum Physics, University of Science and Technology of China, Hefei, 230026, China
\end{affiliations}

\newpage

\begin{abstract}
Quantum mechanics provides a statistical description about nature, and thus would be incomplete if its statistical predictions could not be accounted for some realistic models with hidden variables\cite{einstein_can_1935}. There are, however, two powerful theorems against the hidden-variable theories showing that certain quantum features cannot be reproduced based on two rationale premises of classicality, the Bell theorem\cite{Bell.RevModPhys.38.447.1966}, and noncontextuality, due to Bell, Kochen and Specker (BKS) \cite{Kochen.JMathMech.17.59.1967}. Tests of the Bell inequality and the BKS theorem are both of fundamental interests and of great significance \cite{hensen_loophole_2015,lapkiewicz_2011_nature}. The Bell theorem has already been experimentally verified extensively on many different systems \cite{Pan2000,Marcikic2003,Barrett2004,Hofmann2012,Ansmann2009,Giustina2013}, while the quantum contextuality, which is independent of nonlocality and manifests itself even in a single object, is experimentally more demanding \cite{Guehne2010}. Moreover, the contextuality has been shown to play a critical role to supply the `magic' for quantum computation \cite{howard_contextuality_2014}, making more extensive experimental verifications in potential systems for quantum computing even more stringent. Here we report an experimental verification of quantum contextuality on an individual atomic nuclear spin-1 system in solids under ambient condition. Such a three-level system is indivisible and thus the compatibility loophole, which exists in the experiments performed on bipartite systems, is closed. Our experimental results confirm that the quantum contextuality cannot be explained by nonlocal entanglement, revealing the fundamental quantumness other than locality/nonlocality within the intrinsic spin freedom of a concrete natural atomic solid-state system at room temperature.
\end{abstract}

In quantum mechanics, not all properties can be simultaneously well defined.
Such incompatibility of properties, characterized by Heisenberg uncertainty principle,
is one of the most curious and surprising features of quantum mechanics, providing the ¡®magic¡¯ for quantum computation
and conflicts strongly with our experience in daily lives.
Hidden variable (HV) theory aims at extending quantum mechanics into a more fundamental theory
which provides a classical-like deterministic description of the nature.
An intuitive feature of the classical description is { its non-contextuality}:
the result of a measurement of an observable is predetermined and independent of which set of compatible (i.e., commeasurable) observables might be measured alongside.
Namely, if $A$, $B$ and $C$ are observables such that $A$ and $B$ commute, $A$ and $C$ commute,
but $B$ and $C$ do not commute, then the value predicted to occur in a measurement of $A$
does not depend on whether $B$ or $C$ was measured simultaneously.
The theorem derived by Bell \cite{Bell.RevModPhys.38.447.1966}, Kochen and Specker \cite{Kochen.JMathMech.17.59.1967}, called BKS theorem, show that non-contextuality hidden variable (NCHV) is in conflict with quantum mechanics.

To confirm the theorem, many theoretical schemes \cite{Mermin_1990,Mermin_1993,peres_incompatible_1990,Cabello.PhysRevLett.80.1797.1998,Simon.PhysRevLett.86.4427.2001,Larsson.EPL.58.799.2002,Spekkens.PhysRevA.71.052108.2005} have been proposed for possible experimental tests of quantum contextuality. Unfortunately, it had been a conundrum for experimentalists to accomplish such a test, because the BKS theorem was generally nullified in real experiments due to the more demanding measurement precision \cite{Meyer_1999_PhysRevLett.83.3751}. Not until recently it was shown that the BKS theorem could be converted into experimentally available schemes by correlations between compatible measurements based on some inequalities called non-contextuality inequalities \cite{Cabello_2008_PhysRevLett.101.210401,Badzia_2009_PRL_050401,Yu_2012_PRL_030402}.

Several experimental tests of the BKS theorem were performed using single photons \cite{michler_experiments_2000,huang_experimental_2003,Amselem_2009}, neutrons \cite{Hasegawa2006,Bartosik2009}, ions \cite{kirchmair_state-independent_2009} and NMR \cite{moussa_testing_2010} systems. Although the results obtained in those experiments were completely in conflict with non-contextuality, the involvement of at least two particles in those experiments left some loopholes open, such as the uncontrollable inter-particle interactions probably reducing the compatibility of the measured observables, and the detection loophole in multi-photon experiments due to photon loss and phase instability. In this sense, experiments performed on individual single-particle systems are more compelling and highly desirable, which are also of much more challenging with currently available technology.

Previous verifications of the BKS theorem for a spin-1 particle \cite{Spekkens.PhysRevA.71.052108.2005,peres_incompatible_1990} required tens of observable quantities. It was an insurmountable obstacle for experimentalists to find a qualified system to measure all those observable quantities with enough precision to fulfil this goal. Although a few more experiments demonstrating the conflict with NCHV theories were implemented using single photons \cite{lapkiewicz_2011_nature,zu_2012_PRL_150401} and ions \cite{zhang_2013_PRL}, experimental verifications using individual quantum atomic solid-state systems are desirable but elusive.

Here we report an experimental realization of a genuine single-particle verification of the quantum contextuality by measuring the five properly chosen observables in a spin-1 system, the $^{14}$N nuclear spin of nitrigen-vacancy (NV) center in Diamand. This has been made possible by the  synergy of quantum control of the nuclear spin and sequential measurement procedure assisted by solid immersion lens in diamond. As an outstanding quantum systems with many applications, the NV system has just been adopted to perform a loophole-free Bell inequality\cite{hensen_loophole_2015,giustina_significant_2015}.

Here we report an experimental realization of a genuine single-particle verification of the quantum contextuality by measuring five properly chosen observables in a spin-1 system. Since it is an unmovable and individual system, we prevent a compatibility loophole \cite{guehne_compatibility_2010} regarding inter-particle interactions or entanglement. Thus all nonlocality is excluded from our test. We show below that our experiment has fulfilled a precise detection of a small violation of the non-contextuality inequality.

To begin with, we briefly outline the theoretical scheme that excludes NCHV models for a spin-1 system. Considering five observables $L_i~(i=1,\cdots, 5)$ taking values in the set $\{0,1\}$.  NCHV predicts\cite{lapkiewicz_2011_nature}:
\begin{eqnarray}\label{NCHV_inequality}
\left ( \sum_{i=1}^{5}\langle L_i\rangle - \sum_{i=1}^{5} \langle L_i L_{i+1}\rangle\right )_{\mathrm{NCHV}} \leqslant 2.
\end{eqnarray}
 with identification $L_6$ equals $L_1$, here the expectation values are taken with respect to certain probability distribution of the hidden variables that determine the values of $L_i^{'}$s. In  actual measurements, the observable $L_1^{\prime} = L_6$ may differ from $L_1$ due to experimental imperfection. Hence the inequality (1) should be modified \cite{lapkiewicz_2011_nature,um_experimental_2013}  as
\begin{equation}
\left ( \sum^{5}_{i=1}\langle L_i\rangle -\sum^5_{i=1}\langle L_i L_{i+1}\rangle  -\langle L_1 \rangle +\langle L^{\prime}_{1}L_1\rangle \right ) _{\mathrm{NCHV}} \leqslant 2.
\label{mod_NCHV_inequality}
\end{equation}

For a spin-1 system we denote by $S_{\ell_j}^2$ the square of the spin operator along the unit vector $\ell_j$.
If two unit vectors $\ell_i$ and $\ell_j$ are orthogonal, then the observables $S_{\ell_i}^2$ and $S_{\ell_j}^2$ are compatible and can be measured simultaneously.
We consider a cyclic quintuplet of unit vector with $\ell_i\perp\ell_{i+1}$
($i= 1, \ldots,6$ and $\ell_6=\ell_1$, see details in SI) and five
corresponding observables $L_i=$I$-S_{\ell_i}^2=\ket{\ell_i}\bra{\ell_i}$ where the observable equals 1 if spin is parallel with the unit vector and equals 0 if spin neutral or anti-parallel with the unit vector. The expectations $\langle L_i\rangle$ are evaluated with respect to a certain state $\ket{\psi_0}$ of the particle. We project our nuclear spin into neutrally polarized state $\ket{\psi_0}$ directed along the fivefold symmetry axis of the regular pentagram, the vertices of which correspond to the five vectors $\ell_i$. From quantum mechanics prediction it follows from the facts $\langle L_i\rangle_{\psi_0} = \bra{\psi_0}L_i\ket{\psi_0}=|\bra{\psi_0}\ell_i\rangle|^2=1/\sqrt{5}$.  And due to $\langle L_i L_{i+1}\rangle_{\psi_0} =  \bra{\psi_0}\ell_i\rangle \bra{\ell_i}\ell_{i+1}\rangle \bra{\ell_{i+1}} \psi_0\rangle = 0$, the inequality will follow as that:
\begin{eqnarray}\label{quantum inequality}
 \sum_{i=1}^{5}\langle L_i\rangle_{\psi_0} - \sum_{i=1}^{5} \langle L_i L_{i+1}\rangle_{\psi_0}   =\sqrt{5} > 2.
\end{eqnarray}
This means that quantum mechanical prediction violates for state $\psi_0$ the NCHV inequality (1).

\begin{figure}
\centering
\includegraphics[width=1\columnwidth]{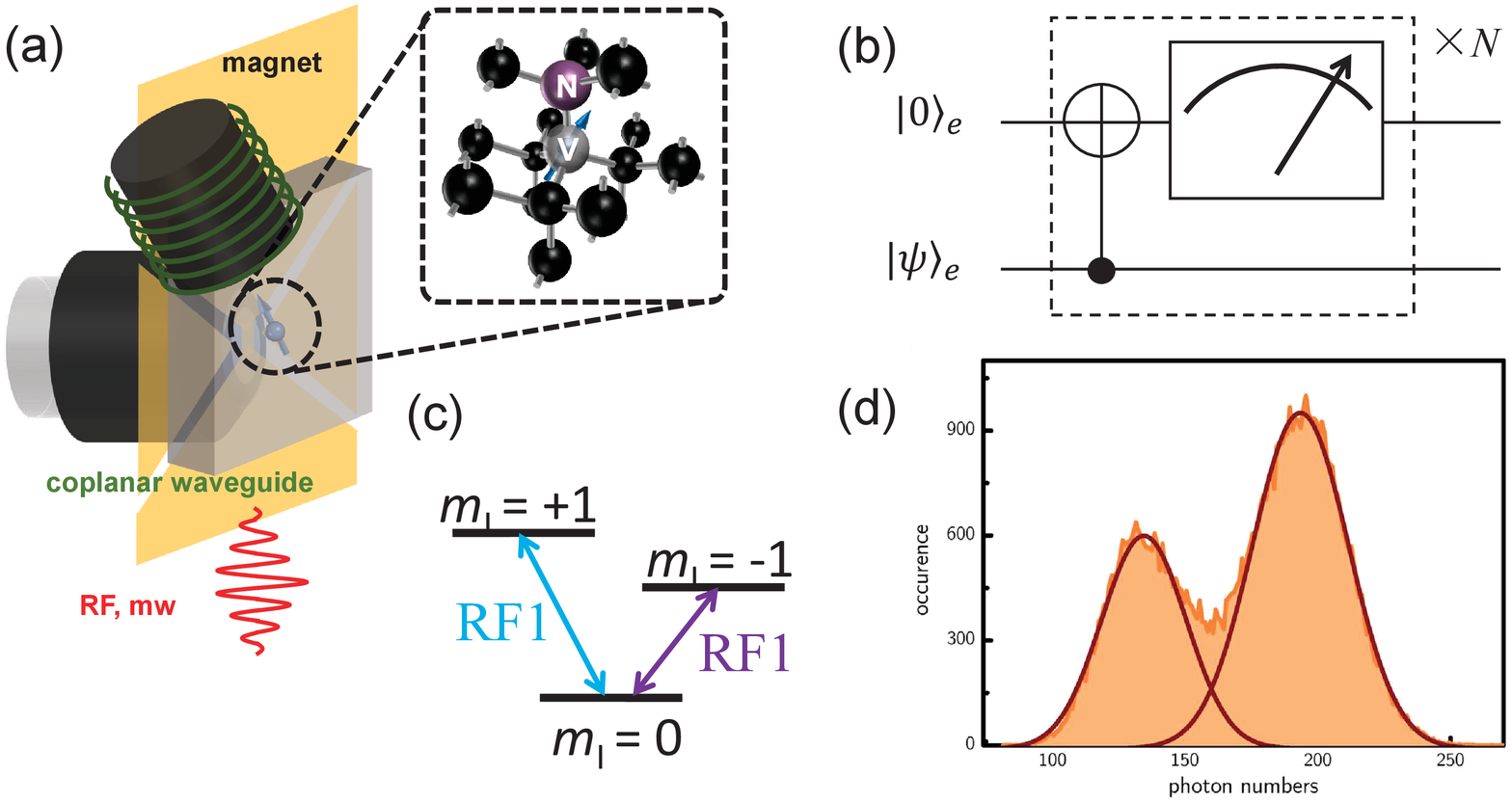}
\caption{ Single intrinsic qutrit spin system. (a) Schematic the confocal microscopy setup. The state of qutrit was initialized and readout by green laser illumination through the objective. An external DC magnetic field is applied by a movable permanent magnet while microwave and RF is carried by coplanar waveguide. (b) Non-destructive readout scheme of qutrit \cite{neumann_science_2010}. (c)  The nuclear energy level scheme of the nitrogen nuclear spin of NV defect, where the left and right lines denote the transitions for the lower frequency RF pulse (RF1) and the higher frequency RF pulse (RF2), respectively. (d) Occurrence of photon numbers histogram of single shot readout of nuclear spin.
}
 \label{fig1}
\end{figure}

In this experiment we will show the violation of the prediction of a general NCHV model. We employed an intrinsic atomic solid-state qutrit at room temperature, a single spin-1 nitrogen nuclear spin in a negatively charged nitrogen-vacancy (NV) center in diamond. A single electron spin is adjacent to the qutrit and served as an ancilla to initialize and readout the nuclear spin qutrit. With weak interaction, high magnetic field and high fluorescence collection efficiency (assisted by solid immersion lens (SIL) \cite{hadden_strongly_2010}),  a non-destructive projective single-shot readout of the nuclear spin qutrit state is fulfilled by mapping the state onto the electron spin, which can then be read out repeatedly \cite{neumann_science_2010}. The Hamiltonian of the nitrogen nuclear spin is given by H$=Q \hat{I}^2_z+\gamma_n\mu_nB_z\hat{I}_z$ , where the zero field splitting $Q$=4.95~MHz and the nuclear gyromagetic ratio $\gamma_n = 0.3077$~kHz/G. A 5636~G magnetic field is applied along the crystal axis $ \langle 111 \rangle$, and the energy splitting depends linearly on the magnitude of external magnetic field. Individual NV center is optically addressed by a home-built confocal microscopy. Fig.~\ref{fig1}~(a,~b) show the structure of the qutrit and the scan map. The qutrit energy levels is shown in Fig.~\ref{fig1}~(c), as revealed by the NMR spectrum of nitrogen nuclear spin [Fig.~\ref{fig1}~(d)]. The required dual-channels resonant RF pulses are applied to the nuclear spin for state manipulation.  Spin-state-dependent photon number histogram is shown in Fig.~\ref{fig1}~(d). An optimized threshold is set to distinguish between $|1\rangle$ and $|0,-1\rangle$ state. In the observation procedure of this experiment, the nuclear spin is firstly initialized to $|1\rangle$ state (the initialized threshold is set to be lower to get high initialization fidelity\cite{neumann_science_2010}). Then another 593~nm laser is applied for 10 ms to distinguish the charge state of NV center \cite{waldherr_violation_2011}. The NV charge state will not be affected by nuclear spin state single shot readout procedure (see SI for detail). Other operators are realized by applying different unitary transformations that map the observable of interest onto the readout basis states.


The five states $l_i ~(i = 1,\ldots,5)$ involved in the measurements are selected as
\begin{equation}\label{basis}
\begin{aligned}
 &\left| {{l_1}} \right\rangle  = \left| { + 1} \right\rangle ,\left| {{l_2}} \right\rangle  = \left| { - 1} \right\rangle ,\left| {{l_3}} \right\rangle  = {R_a}( - \gamma )\left| {{l_1}} \right\rangle , \\
 &\left| {{l_4}} \right\rangle  = {R_a}( - \gamma ){R_b}( - \gamma )\left| {{l_2}} \right\rangle ,
 \left| {{l_5}} \right\rangle  = {R_a}( - \gamma ){R_b}( - \gamma )\left| {{l_3}} \right\rangle , \\
 &\left| {{l_6}} \right\rangle  = {R_a}( - \gamma ){R_b}( - \gamma )\left| {{l_4}} \right\rangle  = \left| {{l_1}} \right\rangle , \\
 \end{aligned}
\end{equation}
with the angle $\gamma  = \arccos \left( {2 - \sqrt 5 } \right)$ ( the rotation operators $R_a(\theta)$ and $R_b(\theta)$ are present in detail in SI).
The maximal violation is attained by the quantum state $\left| {{\psi _0}} \right\rangle  = \frac{1}{{\sqrt[4]{5}}}\left| { + 1} \right\rangle  + \sqrt {1 - \frac{2}{{\sqrt 5 }}} \left| 0 \right\rangle  + \frac{1}{{\sqrt[4]{5}}}\left| { - 1} \right\rangle  = {R_a}(\phi ){R_b}( - \theta ){R_a}( - \pi )\left| { + 1} \right\rangle$,
with the angles $\theta  = \arccos \left( {1 - \frac{2}{{\sqrt 5 }}} \right)$ and $\phi  = \arccos \left( {\frac{{1 - \sqrt 5 }}{2}} \right)$.
The experimental schematic procedure is shown in Fig.~2~(b), while the detail pulse sequence is shown in Fig.~2~(c).
At first, the specific state $\left| {{\psi _0}} \right\rangle$ is prepared from the initial state $\left| { +1} \right\rangle$ by ${U_{{\rm{ini}}}} = {R_a}(\phi ){R_b}( - \theta ){R_a}( - \pi )$ (the 1st and 2nd box in Fig.~2~(b)). In such a way, the spin state is made into neutrally polarized state with respect to fivefold symmetry axis of the regular pentagram.
Next, one of the five unitary operations ${U_1} = I,{U_2} = {R_a}(\gamma ),{U_3} = {R_a}(\gamma ){R_b}(\gamma ),{U_4} = {R_a}(\gamma ){R_b}(\gamma ){R_a}(\gamma ),$ and ${U_5} = {R_a}(\gamma ){R_b}(\gamma ){R_a}(\gamma ){R_b}(\gamma )$ is applied ( the 3rd box in Fig.~2~(b), see SI for details). Such a transformation will map the measurement basis into \{$\ell_i$, $\ell_{i+1}$, and $\ell_i \times \ell_{i+1}$\}.
 The combined effect of ${U_i} ~(i=1,\ldots,5)$ and the measurement on $\left| { +1} \right\rangle$ amounts to the measurement on $U_i^\dag \left| { + 1} \right\rangle  = \left| {{l_{2\left\lfloor {i/2} \right\rfloor  + 1}}} \right\rangle$, where $\left\lfloor  \right\rfloor$ denotes the floor function. After that, a single-shot readout of the nuclear spin state obtains the probability on $\left| { +1} \right\rangle$ (the first readout of measurement procedure in Fig.~2~(c)).
Afterwards, the probability on $\left| { +1} \right\rangle$ and $\left| { -1} \right\rangle$ is swapped by an operation ${U_{{\rm{swap}}}} = {R_b}(\pi ){R_a}(\pi ){R_b}(\pi )$, and then the probability on $\left| { +1} \right\rangle$ extracted by another single-shot readout (the second readout of measurement procedure in Fig.~2~(c)). The combined effect of ${U_i}$, ${U_{{\rm{swap}}}}$, and the subsequent measurement on $\left| { +1} \right\rangle$ amounts to the measurement on $U_i^\dag \left| { - 1} \right\rangle  = \left| {{l_{2\left\lfloor {(i + 1)/2} \right\rfloor }}} \right\rangle$.
Due to the high readout fidelity, the experimental error incurred by the above sequential measurement is less than 5\%. Therefore, the expectation of the operator ${L_i}{L_{i + 1}}$  is revealed. Note that in reality, $L_6$  differs from $L_1$ due to experimental imperfections, and Eq. (2) rather than Eq. (3) is to be falsified ( the pulse sequence to measure $\langle L_1\rangle$, $\langle L'_1L_1\rangle $ is shown in Fig.~2~(d)).

The measurement of inequality (2) with $\ket{\psi_0}$ for all the involved observables are shown in FIG.~\ref{fig3}. The experimental results of the above gives:
\begin{eqnarray}\label{exp inequality}
  \sum_{i=1}^{5}\langle L_i\rangle_{\psi} - \sum_{i=1}^{5} \langle L_i L_{i+1}\rangle_{\psi}  -\langle L_1\rangle+\langle L'_1L_1\rangle \nonumber
  \\ =  2.117~(\pm 0.015)
\end{eqnarray},
 with the value of each term is plotted in Fig.\ref{fig3}.
This result shows  the violation of the NCHV inequality (2) by about 7.8 ($0.117$ over $0.015$) times of the standard deviation, as expected by quantum mechanical expectation given by (3).

In summary, our experimental results have demonstrated unambiguously the violation of the non-contextuality inequality, i.e., the results cannot be explained by any non-contextual model. Remarkably, our implementation, performed on a single particle in atomic solid state system, has closed the compatible loophole which exists in the experiments performed on bipartite systems, and should be more convincing. Our experiment may give profound impacts on quantum mechanical foundation, and we expect this result to stimulate more elegant experiments for further exploration of the peculiar characteristic and profound implications of quantum physics.


\begin{addendum}
 \item This work was supported by the National Key Basic Research Program of China (Grant
No. 2013CB921800), the National Natural Science Foundation of China (Grant No. 11227901, No. 31470835, No. 11274299, No. 21303175, No. 21322305, No. 21233007, and
No. 11275183), and the Strategic Priority Research Program (B) of the CAS (Grant No. XDB01030400). F. S.
and X. R. thank support of the Youth Innovation Promotion Association of Chinese Academy of Sciences.
 \item[Competing Interests] The authors declare that they have no competing financial interests.
 \item[Author contributions] J.D., M.S. and X.K. proposed the idea and designed the experimental proposal. X.K., F.S, P.W., prepared the experimental set-up. X.K., F.S, P.W., P.H., F.K. performed the experiments. M.S, X.K, S.Y, C.D and W.M. carried out the theoretical calculation. M.W., H.D. and L.M. prapare the SIL. J.D. supervised the experiments.
 X.K., C.D. and J.D. wrote the paper. All authors analyzed the data, discussed the results and commented on the manuscript.
 \item[Correspondence] Correspondence and requests for materials should be addressed to J.D.\ (email: djf@ustc.edu.cn).
\end{addendum}




\begin{figure}
\centering
\includegraphics[width=1\columnwidth]{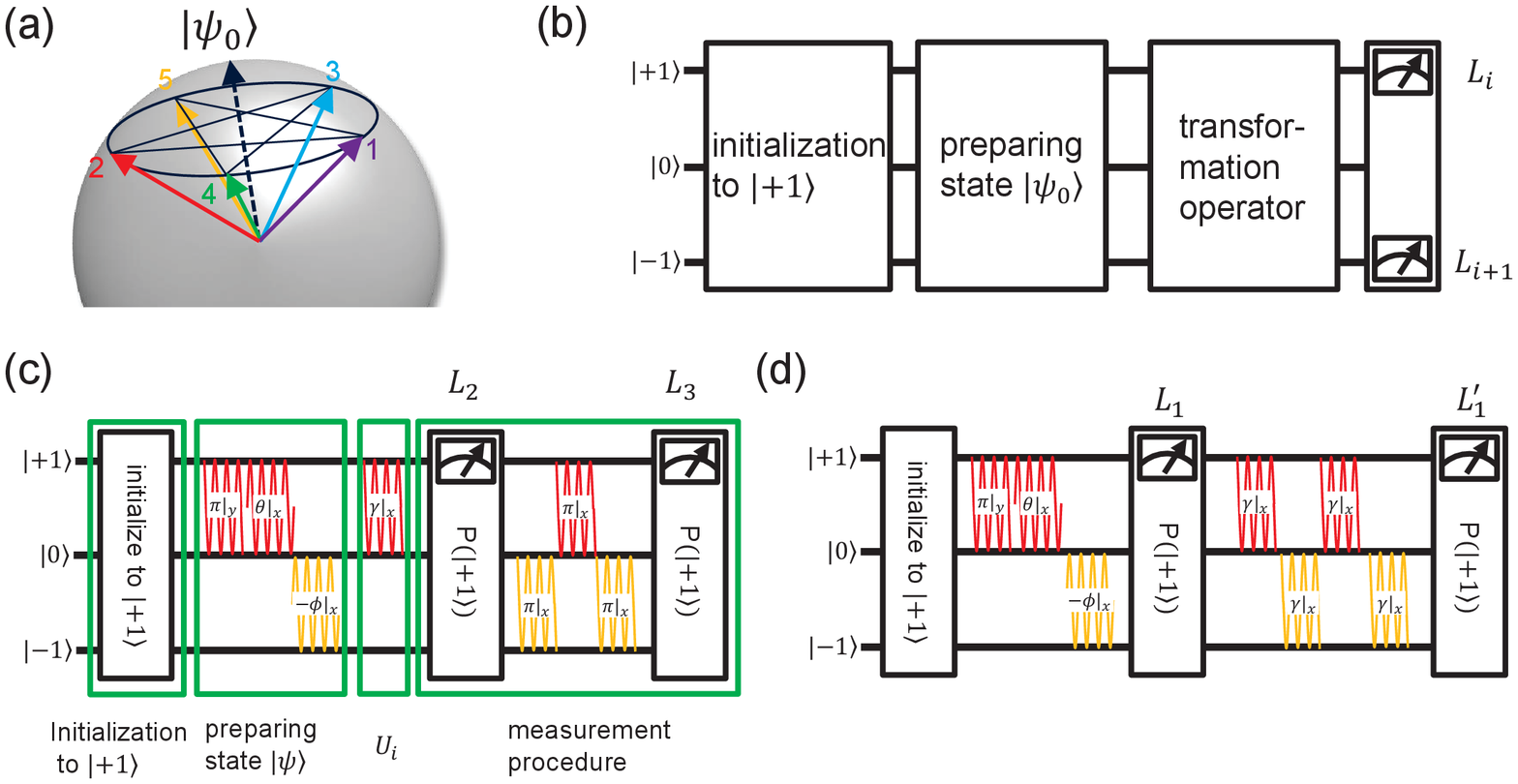}
\caption{(Color online). {Pulse sequences for test of noncontextuality measurements.}  (a) Five regular pentagram vectors (b) Block diagram for the measurements of two adjacent compatible operators In the first stage, the initial state $\ket{\psi_0}$ is implemented with RF sequence  $R_{2}(-\phi)R_{1}(\theta)R_{1y}(\pi)$. Secondly, unitary operation $U_2$=$R_1(\gamma)$ is applied to map $|-1\rangle\langle -1|$, $|1\rangle\langle1|$ to observables $|\ell_2\rangle\langle\ell_2|$ and $|\ell_3\rangle\langle\ell_3|$.   (c) A single-shot readout of nuclear spin state reveal the observable as $L_2=P(|1\rangle)$ . A temporal sequential measurement{ \cite{kirchmair_state-independent_2009}} is implemented to get the correlation terms $\langle L_2L_{3}\rangle$=$P(L_2$=$1,L_{3}$=$1)$.  The RF pulse sequence $R_1(\pi)R_2(\pi)R_1(\pi)$ is applied to measure another observable with the value $P(|-1\rangle)$. We show here the experimental sequence of $\langle \ell_2 \rangle$ and $\langle \ell_2\ell_3 \rangle$, others are similar to this procedure. (d) The sequence is applied to $\langle L_1\rangle-\langle L'_1L_1\rangle $, $L'_1=L_6$.}
 \label{fig2}
\end{figure}

\begin{figure}
\centering
\includegraphics[width=0.8\columnwidth]{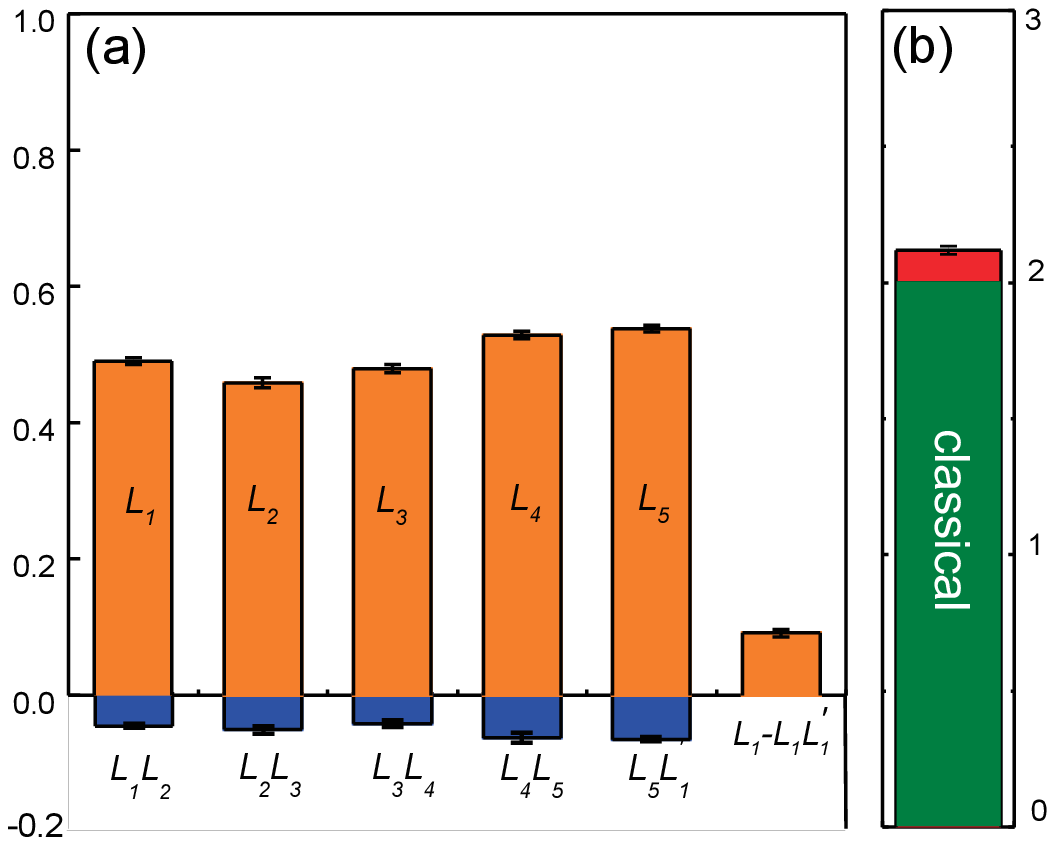}
\caption{(Color online). Experimental results for the quantum contextuality verification. (a) Experimental measurements for the left hand side of Eq.~(\ref{mod_NCHV_inequality}). The measurement results for $L_i(i=1,...,5)$ shown in orange bar. And the correlation terms $L_iL_{i+1}(i=1,...,5)$ are shown in blue bar. (b) The final result for left term of Eq.~(\ref{exp inequality}). Green bar shows the classical prediction, and the red bar shows the exceeding of experimental results.
\label{fig3}
}
\end{figure}





\end{document}